\documentclass{article}
\usepackage{graphicx}
\usepackage{amssymb}
\newcommand{\bm}[1]{\mbox{\boldmath $#1$}}     
\newcommand{\ec}[2]{\begin{equation}\label{#1} #2\end{equation}}
\newcommand{\dist}{\mbox{dist\,}}
\newcommand{\diag}{\mbox{diag\,}}

\DeclareMathAlphabet{\mathsfsl}{OT1}{cmss}{m}{sl}
\DeclareMathAlphabet{\mathbsl}{OT1}{cmr}{bx}{sl}
\DeclareMathAlphabet{\mathbit}{OT1}{cmr}{bx}{it}
\DeclareMathAlphabet{\mathb}{OT1}{cmr}{bx}{rm}

\begin{document}

    \title{Geometry of an accelerated rotating disk}
    \author{J.-F. Pascual-S\'anchez, A. San Miguel, F.
    Vicente \\[2mm]
    {\normalsize Dept. Matem\'atica Aplicada, Facultad de Ciencias} \\
     {\normalsize Universidad de Valladolid, 47005, Valladolid }\\
     {\normalsize Spain}}

\date{}
\maketitle

\begin{abstract}
We analyze the geometry of a rotating disk with a tangential
acceleration in the framework of the theory of  Special
Relativity, using the kinematic linear differential system that
verifies the relative position vector of time-like curves in a
Fermi reference. A numerical integration of these equations for a
generic initial value problem is made up and the results are
compared with those obtained in other works.
\end{abstract}


\section{Introduction}
The geometry of a rotating disk has generated an enormous
literature (see the recent review by Gr\o n~\cite{ref:Gr03} in
which one of the basic problems treated is connected with the
precise definition of the space representing the disk. In
applications such as the study of the motion of a rotating disk,
when gravitational effects are discarded, splits of the Minkowski
spacetime naturally occurs. One of these splits considers the
Minkowski spacetime together with a congruence defined by a
timelike vector field. Then, following Cattaneo~\cite{ref:Ca58}, a
{\it spatial metric} $ds^\perp$ and a {\it standard time} $dx^0$
relative to the congruence can be introduced, so that quantities
such as the ratio $ds^\perp/dx^0$ have a physical and operational
meaning. The relationship  of this congruence splitting and that
defined by a foliation of spacetime by spacelike hypersurfaces has
been  shown for general curved spacetimes, e.g., in  Bini {\it et
al.}~\cite{ref:JC89}.  The ``hypersurface point of view'' for the
rotating disk has been considered , among others, by Tartaglia
~\cite{tar} for a definition of ``space'' of a rotating disk and
by the present authors, in a discussion of the Sagnac effect
~\cite{PSV03}. The ``congruence point of view'' has been recently
used in the work by Rizzi \& Ruggiero~\cite{ref:RiRu02}, for the
study of the space geometry of rotating platforms and, by
Minguzzi~\cite{ref:Mi03}, for the study of simultaneity in
stationary extended frames.

In \cite{ref:RiRu02}, it is shown that each element of the
periphery of the disk, of a given proper length, is stretched
during the acceleration period using for this the Gr\o n model
\cite{ref:Gr79}, in which the motion of the disk is not Born-rigid
in the acceleration period. In \cite{ref:Gr79}, this dilatation of
length is discussed using a kinematical argument to calculate this
change taking into account the asynchrony of the acceleration
measured from the rotating frame.

In this work we study the evolution of the distance between any
two nearby points on the periphery of an accelerated rotating
disk, using the geometrical properties of the timelike congruences
in the special relativistic spacetime associated to the disk. The
motion is described from a Fermi reference field, solving
numerically the differential equations of the  separation or
relative position vector of timelike curves. Note that we do not
introduce any coordinate system in a reference field as in
\cite{mas}.

The paper is organized as follows. In Sec. 2 it is described the
material system showing the properties of the congruences
associated to a uniformly accelerated rotating disk. In Sec. 3 we
construct a field of Fermi references on the world tube
corresponding to the flow of the disk in the spacetime. Further,
in this section, the evolution of the length of an arc of
circumference is established as an initial value problem for the
deviation  of nearby timelike curves. In Sec. 4 this problem is
numerically integrated and the results obtained are compared with
those given by Gr\o n.

\section{Description of the material system}
Let us consider the usual system of Cartesian coordinates
$\mathrm{x}=(x,y,z,t)$ defined on the Minkowskian space-time
$\mathcal{M}$, and assume that at time $t=0$ the coordinate origin
$O$ is on the center of a circular disk whose radius is $R$ and
its symmetry axis is $z$. Throughout this paper we restrict
ourselves to the three-dimensional submanifold $\mathcal{T}\subset
\mathcal{M}$ given by $z=0$. The $\mathbf{\eta}$--orthonormal
reference field
$\bm{\mathrm{e}}_{\mathrm{a}}:=\{\partial_t,\partial_x,\partial_y\}$
is defined on $\mathcal{T}$. Due to the cylindrical symmetry of
the problem it will be useful to introduce a cylindrical
coordinate system on $\mathcal{T}$ with origin at $O$, defined by
    \ec{310031}{t=t,\quad x=r\cos\phi,\quad y=r\cos\phi.}
These coordinate systems determine a reference frame
$\bm{e}_a:=\{\partial_t, \partial_r,\partial_\phi\}$ on
$\mathcal{M}$, related to $\{\bm{\mathrm{e}}_{\mathrm{a}}\}$ by
the classical expressions
    \ec{310032}{\bm{e}_0=\bm{\mathrm{e}}_0,\quad \bm{e}_1=r\cos\phi\,
    \bm{\mathrm{e}}_1+r\sin\phi\, \bm{\mathrm{e}}_2,\quad
    \bm{e}_2=-r\sin\phi\, \bm{\mathrm{e}}_1+r\cos\phi\, \bm{\mathrm{e}}_2.}

In addition to the coordinate systems $\{\bm{\mathrm{x}}\}$ and
$\{\bm{x}\}$, we also use a convected coordinate system:
$\{X\}:(T,R,\Phi)$, co-rotating with the disk, defined in terms of
$\{\bm{x}\}$ by the coordinate transformation:
    \ec{310033}{T=t,\quad R=r, \quad \Phi=\phi-\phi(t),}
where $\phi(t)$ is a smooth function of $t$. The corresponding
reference frame field on $\mathcal{M}$ is given by
$\bm{E}_A:=\{\partial_T,\partial_R,\partial_\Phi\}$, which is
related to $\{\bm{e}_a\}$ through the expressions
    \ec{310034}{\bm{E}_0=\bm{e}_0+\dot{\phi}(t)\bm{e}_2, \quad \bm{E}_1=
    \bm{e}_1,\quad \bm{E}_2=\bm{e}_2.}

In the Cartesian coordinates $\bm{\mathrm{x}}$, the matrix
representation of the Min\-kows\-kian metric is
    $\bm{\eta}(\bm{\mathrm{e}}_{\mathrm{a}},
    \bm{\mathrm{e}}_{\mathrm{b}})=\eta_{\mathrm{ab}}$,
where $\eta_{\mathrm{ab}}=\diag(-1,1,1)$. In terms of convective
coordinates $\{\bm{X}\}$ this metric takes the form
    $\bm{\eta}(\bm{E}_{A},
    \bm{E}_{A})=G_{AB}$,
where  $G$ is the matrix
    \ec{710031}{G = \left(\begin{array}{ccc}
    R^2\varpi(T)^2 -1 & 0 & R^2\varpi(T) \\
    0 & 1 &  0 \\
    R^2\varpi(T) & 0 & R^2
    \end{array}\right)}
where $\varpi(T):=\omega+T\alpha$, $\omega:=\dot{\phi}(t)$ is the
coordinate angular speed  and $\alpha:= \ddot{\phi}(t)$ the
coordinate angular acceleration.

In the coordinate system  $\{\bm{\mathrm{x}}\}$ fixed in the
space-time, the worldlines of points in the accelerated rotating
disk are curves parametrized by the coordinate time
    \ec{111034}{\bm{\mathrm{x}}(t)  =\bigg(t, \, r_0\cos\big(\phi_0+\phi(t)\big),\,
    r_0\sin\big(\phi_0+\phi(t)\big)\bigg).}
Using convective coordinates these curves may be represented as
$\bm{X}(T) = (T, R_0,\Phi_0)$, so that each worldline is
identified by means of a pair $(R_0,\Phi_0)$. The tangent vector
field to this flow is given in convective representation as
    \ec{111031}{  \dot{\bm{X}}(T) = \bm{E}_0,}
where $\dot{\bm{X}}$ denotes the derivative of $X$ with respect to
$T$. Therefore, the tangent vector field to the flow, given by the
space-time velocity
\ec{710032}{\bm{V}:=(-G_{00}(X))^{-1/2}\bm{E}_0,}
takes in the
convective representation the expression
    \ec{710032}{\bm{V} = \big((1-R^2\varpi(T)^2)^{-1/2},0, 0\big).}

For the vector field $\bm{V}(T)$ on the curve $\bm{X}(T)$ the
proper acceleration $A^I$  in convective coordinates is
    \ec{310039}{A^I:=\frac{DV^I}{ds} =\frac{dV^I}{ds}+\Gamma^I_{BC}V^B\frac{dX^C}{ds},}
where $s$ denotes the proper time on the curve $\bm{X}(T)$.

\section{Construction of a Fermi reference frame field}
Now we will construct on a curve $\bm{X}(T)=(T,R,\Phi)$ a
$G$--orthonormal reference frame field satisfying the Fermi
propagation law. Given a $G$--orthonormal reference,
$\bm{E}_{\hat{a}}(0)$, at an initial coordinate time $T=0$, with
$\bm{E}_{\hat{0}}(0)=\bm{V}(0)$, the relation between this
reference and the comoving coordinate basis is
    \ec{3100310}{\bm{E}_{\hat{a}}(0)=E_{\hat{a}}^A(0)\, \bm{E}_A(X_0).}
In the Fermi transport the evolution of the vector
$\bm{E}_{\hat{0}}(0)$ is given by $\bm{V}(s)$. Thus, only the
evolution of the spacelike vectors $\bm{E}_{\hat{\alpha}}$ of the
reference frame must be determined. The absolute derivative of the
field $\bm{E}_{\hat{\alpha}}$, satisfies the equations
    \ec{610031}{\frac{DE^I_{\hat{\alpha}}}{ds} = V^I A^B G_{BC}
    E^C_{\hat{\alpha}},
    \qquad\qquad\hat{\alpha}=1,2,}
which can be expressed in terms of the coordinate time as the
ordinary differential equation
    \ec{610034}{\frac{dE^A_{\hat{\alpha}}}{dT} = \sqrt{-G_{00}}(- \Gamma^A_{BC}V^B +
    V^AA^BG_{BC})E^C_{\hat{\alpha}}.}
The integration of this differential system with the initial
condition (\ref{3100310}) gives the evolution of
$\bm{E}_{\hat{a}}(T)=E^A_{\hat{a}}(T)\,\bm{E}_A(T)$, i.e., of the
Fermi reference along the curve $\bm{X}(T)$.

The circumference of the rotating disk is identified with any of
the circumferences $\mathcal{C}_T$ obtained by cutting the world
tube of the disk with planes $t={\rm const}$ which are orthogonal
to the worldline of the center of the disk. The construction of
the quotient space $\mathcal{D}$ of $\mathcal{T}$ by  the flow of
time-like curves corresponding to  $\mathcal{T}$ determines the
material space associated to the disk. At each time $T$, a value
of the line element is assigned to each circumference
$\mathcal{C}_T$, using the metric induced by the Minkowskian
metric on $\mathcal{T}$.

Let us denote by $C$ the circumference of $\mathcal{D}$ and let
$\tilde{P}$ and $\tilde{P}^\prime$ be two infinitesimally
neighboring points on $C$. Now, consider the timelike curves
$\sigma(s)$ and $\sigma^\prime(s)$ on $\mathcal{M}$ which are
projected on the points $\tilde{P}$ and $\tilde{P}^\prime$
respectively. In an arbitrary point $P$ in $\sigma(s)$, the
tangent space $T_{P}\mathcal{M}$ can be split as
$T_{P}\mathcal{M}=\mathcal{H}_{P}\oplus\mathcal{V}_{P}$, where
$\mathcal{V}_{P}$ represents the vector subspace whose elements
are vectors parallel to the tangent vector $\bm{V}$ to $\sigma(s)$
at the point $P$, and $\mathcal{H}_{P}$ is the $G$--orthogonal
complement to $\mathcal{V}_{P}$. Every vector $\bm{X}\in
T_{P}\mathcal{M}$ can be projected on $\mathcal{H}_{P}$ using the
projector $\bm{h}=\bm{1}+\bm{V}\otimes \bm{V}^\flat$, where
$\bm{V}^\flat$ denotes the 1--form dual to $\bm{V}$. This
projector corresponds to a metric
$\bm{h}^\flat=\bm{G}+\bm{V}^\flat\otimes \bm{V}^\flat$ on
$\mathcal{H}$. In each time $s$, the subspace $\mathcal{H}_P$ may
be identified with the space  $T_{\tilde{P}}\mathcal{D}$. A
further identification may be established between the space-like
vectors  of the Fermi reference $\{\bm{E}_{\hat{\alpha}}(P;s)\}$

and the spacelike vectors of the co-rotating reference
$\{\bm{E}_{\Xi}(\tilde{P})\}$ on $T_{\tilde{P}}\mathcal{D}$.

The measure of the distance between two $P_1\in\sigma_1(s)$ and
$P_2\in\sigma_2(s)$ from a Fermi reference is carried out
determining  the length of the vector  $\bm{S}$ of relative
separation using the metric $\eta_{\hat{a}\hat{b}}=\diag(-1,1,1)$.
On the other hand, in the material description on the quotient
space $\mathcal{D}$, the relative position  of the points
$\tilde{P}_1$ and $\tilde{P}_1$ is constant, however the metric
$\bm{h}(s)$ depends on time.

Following \cite{ref:HaEl73}, we determine the rate of change of
the separation of the points $P_1$ and $P_2$ as measured in
$\mathcal{H}_P$, i.e.,  the rate of change of the relative
position  vector. Let $\lambda(\Phi)$ be a parametrization of the
circumference $\mathcal{C}_0$ at the time $T=0$. The tangent
vector field to $\mathcal{C}_0$ is
$\bm{S}=\partial_\Phi|_{\lambda(\Phi)}$. Consider the family of
curves $\lambda(\Phi,s)$, obtained moving each point
$\lambda(\Phi)$ a distance $s$ on the corresponding integral curve
of the flow of $\bm{V}$. Now, defining the connecting vector field
$\bm{S}:=\partial_\Phi|_{\lambda(\Phi,s)}$ such that the Lie
derivative $\mathcal{L}_{\mathbit{V}}\bm{S}$ is zero, one obtains
that the convective representation of $\bm{S}$ must satisfy
    \ec{410031}{\frac{DS^A}{ds}=V^A_{\;\; ;B}\,S^B.}
Next, defining the separation  or relative position  vector
between the points $\tilde{P}_1,\tilde{P}_2\in\mathcal{D}$,
measured in the convective frame, as
    \ec{1311032}{Y^A:=h^A_{\;\; B}S^B,}
one obtains that the evolution of this vector given by
(\ref{410031}) is $\big(DY^A/ds\big)^\perp=V^A_{\;\; ;B}\, Y^B$,
where $\perp:T_P\mathcal{M}\rightarrow \mathcal{H}$,
$\bm{X}^\perp:=\bm{h}\cdot\bm{X}$. Then, using the definition of
the Fermi derivative of a vector field $\bm{X}$ along a curve
$\sigma(s)$:
    \ec{1211031}{\frac{D_FX^A}{ds}:=\frac{DX^A}{ds}-(X^BA_B)V^A+(X^BV_B)V^A,}
for which the relation $(D\bm{X}^\perp/ds)^\perp=D_F\bm{X}^\perp
/ds$ holds, one can verify that (\ref{1211031}) is equivalent to
$D_FY^A/ds=V^A_{\;\; ;B}Y ^B$. Now, choosing a Fermi reference on
the base curve $\sigma_1(s)$, (\ref{410031}) can be written as
    \ec{510031}{\frac{dY^{\hat{\alpha}}}{ds}=V^{\hat{\alpha}}_{\;\;
    ;\hat{\beta}}Y^{\hat{\beta}},}
where $V^{\hat{\alpha}}_{\;\;;\hat{\beta}}$ is the projection of
the covariant derivative $\nabla \bm{V}$ on $\mathcal{H}_P\otimes
\mathcal{H}^*_P$ expressed in the Fermi reference. Once determined
the matrix $E_{\hat{a}}^A(s)$ from eqn. (\ref{610034}), one obtain
the relation
    \ec{610032}{V^{\hat{\alpha}}_{\;\; ;\hat{\beta}}=E^{\hat{\alpha}}_{\;\; A}E^B_{\;\;
    \hat{\beta}}V^A_{\;\; ;B},}
where $\eta^{\hat{a}\hat{b}}=\eta_{\hat{a}\hat{b}}$.

For the study of the evolution of the interval between the points
$P_1(T)$ and $P_2(T)$, with the same value of $T$, on neighboring
curves measured in a Fermi reference, the basic kinematic
properties of the congruence of world-lines are the vorticity
$\omega_{\hat{\alpha}\hat{\beta}}:= h_{\hat{\alpha}}^{\hat{\mu}}
h_{\hat{\beta}}^{\hat{\nu}} V_{[\hat{\mu};\hat{\nu}]}$, which
represents the angular velocity of the reference frame $\bm{E}_A$
with respect to the Fermi reference $\bm{E}_{\hat{\alpha}}$, and
the expansion $\theta_{\hat{\alpha}\hat{\beta}}:=
h_{\hat{\alpha}}^{\hat{\mu}} h_{\hat{\beta}}^{\hat{\nu}}
V_{(\hat{\mu};\hat{\nu})}$, which represents the rate of change of
the distance between neighboring world-lines. In the particular
case considered in this work, these quantities read:
    \ec{311031}{\omega_{\hat{\alpha}\hat{\beta}}={\small
    -R \varpi(T)\gamma(T)^3\left(\begin{array}{ccc}
    0 & -1 & 0 \\
    1 & 0 & 0\\
    0 & 0 & 0
    \end{array}\right)} }
\ec{311031_}{
    \theta_{\hat{\alpha}\hat{\beta}}={\small
    -R^4\alpha \varpi(T)\gamma(T)^5\left(\begin{array}{ccc}
    0 & 0 & 0 \\
    0 & 1 & 0\\
    0 & 0 & 0
    \end{array}\right),}}
where $\gamma(T):=(1-R^2\varpi^2)^{-1/2}$.

From both the matrix $V^{\hat{\alpha}}_{\;\;\; ;\hat{\beta}}$
given in (\ref{610032}), and an initial valued
$Y^{\hat{\alpha}}(0)$, it is possible to obtain the separation
vector, $\bm{Y}(s)=Y^{\hat{\alpha}}(s)\; \bm{E}_{\hat{\alpha}}$,
by the integration of (\ref{510031}). Given a value of the
parameter $s$, the distance between two points $\tilde{P}_1,
\tilde{P}_2\in\mathcal{D}$ is the number
    \ec{1311031}{\dist(\tilde{P}_1(s),\tilde{P}_2(s)):=(\eta_{\hat{\alpha}\hat{\beta}}
    Y^{\hat{\alpha}}Y^{\hat{\beta}})^{1/2},}
which in convective representation is equivalent to
$\big(h_{AB}(s)Y^A(s)Y^B(s)\big)^{1/2}$, due to the orthogonality
condition $\bm{V}\perp \bm{Y}$.

\section{Integration of the deviation equation of time-like curves}
In order to calculate the instantaneous deviation
$Y^{\hat{\alpha}}(s)$ of two time-like curves, both corresponding
to points on the periphery of the disk, one must solve the linear
differential equation (\ref{510031}) in the independent variable
$T$, whose coefficient matrix is given by (\ref{610032}). This
matrix depends on the coefficients $E^A_{\hat{\alpha}}$ relating
the convective reference $\bm{E}_A$ and the Fermi reference
$\bm{E}_{\hat{a}}$, which satisfy the differential equation
(\ref{610034}). The explicit form of (\ref{310039}) has been
obtained using the {\tt tensor package} contained in the symbolic
processor {\sc Maple}. Equations (\ref{610034}) and (\ref{510031})
lead to a simultaneous system in coordinate time $T$ of eight
first-order differential equations:\\

\begin{eqnarray}
    \dot{x}_1 &=& R\alpha\gamma(T)^2[(-Rx_4x_8-x_4x_6T+x_3x_7T-
    x_3x_8\alpha T^2+ x_3x_8\alpha^3 T^4 R^2    \nonumber\\
    & &\qquad\qquad\quad -x_3x_7T^3R^2\alpha2)x_2+(-x_5x_3\alpha T^2-
    R^2\alpha^2 T^3x_4x_3\nonumber\\
    & &\qquad\qquad\quad -Rx_4x_5+x_5x_3\alpha^3T^4R^2)x_1]\nonumber\\
    \dot{x}_2 &=& R\alpha\gamma(T)^2[(x_8x_6\alpha^3T^4R^2-Rx_7x_8-R^2
    \alpha^2T^3x_7x_6- x_8x_6\alpha T^2)x_2 \nonumber   \\
    & & \qquad\qquad\quad +(x_4x_6T- x_6x_5\alpha T^2-Rx_7x_5-x_3x_7T\nonumber\\
    & & \qquad\qquad\quad -x_6x_4T^3R^2\alpha^2+ x_6x_5\alpha^3T^4R^2)x_1]\nonumber\\
    \dot{x}_3 &=&  -\alpha TRx_4+R\alpha^2T^2x_5\nonumber\\
    \dot{x}_4 &=&  \frac{\alpha Tx_3}{R}+\alpha x_5\label{1311033}\\
    \dot{x}_5 &=& \gamma(T)^2(-R\alpha^2T^2x_3 + \alpha R^2x_4)\nonumber\\
    \dot{x}_6 &=&  -\alpha TRx_7+R\alpha^2T^2x_8\nonumber\\
    \dot{x}_7 &=&  \frac{\alpha Tx_6}{R}+\alpha x_8\nonumber\\
    \dot{x}_8 &=& \gamma(T)^2(R\alpha^2T^2x_6+\alpha R^2x_7)\nonumber
\end{eqnarray}
for the eight unknowns
    \[x_1=Y^1, x_2=Y^2, x_3=E_0^1,
x_4=E_1^1,x_5=E_2^1,x_6=E_0^2,x_7=E_1^2,x_8=E_2^2\]

The initial values are chosen as follows. Let $\bm{E}_A(0)$ be the
convected reference whose origin is on the initial point
$\bm{X}_1(0)$; the matrix $E^{\;\; \hat{a}}_A(0)$ is obtained
orthonormalizing this basis. On the other hand, to choose an
initial value $Y^{\hat{\alpha}}(0)$, we consider, firstly, a
tangent vector $\bm{S}=\ell_0\bm{E}_2\in
T_{\mathbit{X}_1(0)}\mathcal{T}$, where $\ell_0:=R\Delta\Phi$ is
small with respect to the radius of the circumference. Using the
definition of the separation established in (\ref{1311032}), we
will take as the initial value $\bm{Y}(0)=
\bm{h}(0)\cdot\bm{S}(0)$:
    \ec{111033}{ \bm{Y}(0) =\ell_0
    E^{\hat{\alpha}}_2(0)\bm{E}_{\hat{\alpha}}(0),}
where the relations $\bm{V}=\bm{E}_{\hat{0}}$ and
$\bm{E}^{\hat{0}}(\bm{E}_{\hat{0}})=-1$ have been taken into
account.

The differential system (\ref{1311033}) can be solved using
numerical techniques. Here we use the Runge-Kutta-Fehlberg (see,
e.g., \cite{ref:SB92}) method applied to a numerical model for the
accelerate disk. Here we take $R=10^{-5}$ as the radius of the
disk (we are using geometrical units where $c=1$). We assume that
the disk moves uniformly accelerated from the rest, in $T=0$, to
reach an angular velocity $\omega$ at $T_0=3\cdot 10^{-3}$. All
points of the periphery of the disk move with the same angular
acceleration $\alpha=0.25\cdot 10^{8}$ measured from the reference
$\{\bm{e}_{a}\}$ as in Gr\o n's model. We consider the evolution
of the points with angular coordinates $\Theta_1=0$ and
$\Theta_2=10^{-6}$. As for the initial values for the space-like
vectors of the Fermi reference we take
$\bm{E}_{\hat{1}}(0)=(0,1,0)$ and
$\bm{E}_{\hat{2}}(0)=(0,0,R^{-1})$.

The numerical integration of the differential system
(\ref{1311033}), with the considered initial values, has been
realized using the function {\tt ode45} implemented in {\sc
Matlab}, using tolerances {\tt AbsTol} and {\tt RelTol} equal
$10^{-8}$. The evolution $\bm{Y}(T)$ of the deviation vector
expressed in the Fermi reference $\bm{E}_{\hat{a}}$ is shown in
Fig. 1. This deviation increases with the time in the phase of
uniform tangential acceleration
    \begin{figure}[ht]
\begin{center}
    \includegraphics[scale=.405,angle=0]{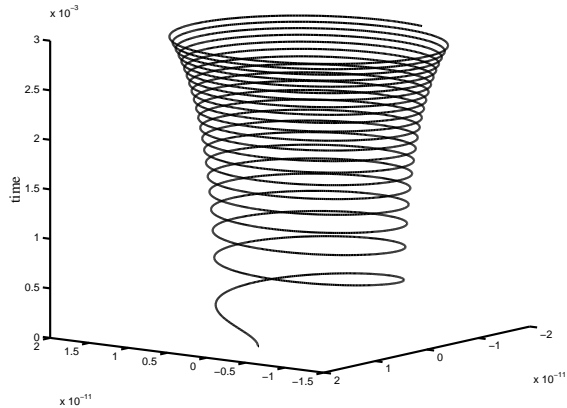}
\end{center}
    \caption{Evolution of the deviation vector between two
    neighboring worldines, represented in a Fermi reference: Horizontal axes
    correspond to the space-like vectors $\bm{E}_{\hat{\alpha}}$ of the reference frame;
    the vertical axis represents the time-like vector $\bm{E}_{\hat{0}}$.}
    \end{figure}

On the other hand, Fig. 2 shows the distance $(\eta_{\hat{\alpha}
\hat{\beta}}Y^{\hat{\alpha}}Y^{\hat{\beta}})^{1/2}$ between points
$P_1(T)$ and $P_2(T)$ as function of time, obtained from the
numerical solution. This figure also shows the corresponding
solution obtained by the Gr\o n method. We observe an complete
agreement between the numerical and  Gr\o n solutions.
    \begin{figure}[ht]
\begin{center}
    \includegraphics[scale=.40 ,angle=0]{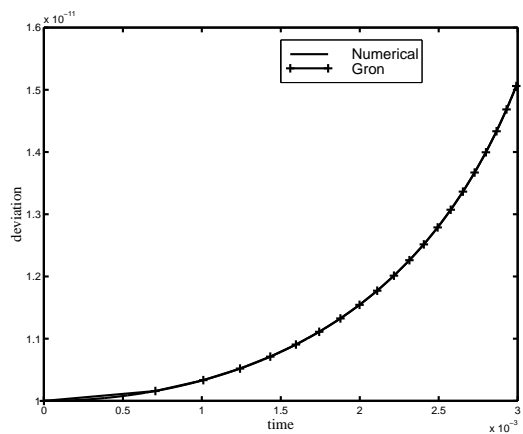}
\end{center}
    \caption{Evolution of the distance between  points on neighboring world-lines.
    The continuous line represents the numerical solution and the line marked
    with ``$+$'' corresponds to the Gr\o n solution.}
    \end{figure}
The method we have studied in this work, may be applied to generic
flows corresponding to others acceleration programs.

\section*{Acknowledgments}  This work was completed with  partial support
from the Junta de Castilla y Le\'on (Spain), project VA014/02.

\end{document}